\documentclass[twocolumn,           
               showpacs,            
               preprintnumbers,     
               aps,                 
               prd,          	    
               a4paper,             
               superscriptaddress,  
               unsortedaddress,     
               footinbib,           
               tightenlines,        
               floats               
               ]{revtex4}

\usepackage{graphicx}
\usepackage{latexsym}
\usepackage{amsmath,amssymb}        
\usepackage[draft=false]{hyperref}

\renewcommand{\[}{\left[}

\font\mybb=msbm10 at 12pt 
\def\bb#1{\hbox{\mybb#1}}

\newcommand{\half}{\ensuremath{\frac{1}{2}}}

\newcommand{\be}{\begin{equation}}
\newcommand{\ee}{\end{equation}}
\newcommand{\ba}{\begin{eqnarray}}
\newcommand{\ea}{\end{eqnarray}}

\begin{document}


\title{On the evolution of cosmic-superstring networks}
\author{Edmund J.~Copeland\footnote{email:Ed.Copeland@nottingham.ac.uk} }
\affiliation{School of physics and astronomy, University of Nottingham,University Park, Nottingham,NG7 2RD.}
\author{P. M. Saffin\footnote{email:p.m.saffin@sussex.ac.uk}}
\affiliation{Department of physics and astronomy, University of Sussex,Falmer,Brighton,BN1 9QJ}

\pacs{98.80.Cq \hfill hep-th/xxxxxxx}
\date{\today} 


\begin{abstract}
We model the behaviour of a network of interacting $(p,q)$ strings from IIB
string theory by considering a field theory containing multiple species of string,
allowing us to study the effect of non-intercommuting events due to two
different species crossing each other. This then has the potential
for a string dominated Universe with the network becoming so tangled that it
freezes.
We give numerical evidence, explained by a one-scale model, that such freezing
does not take place, with the network reaching a scaling limit where
its density relative to the background increases with $N$, the number of string types.
\end{abstract}

\maketitle


\section{Introduction}

Despite the remarkable success of inflation \cite{inflation,review} in describing the observed patterns
of the microwave anisotropy \cite{wmap} there remains an active interest in cosmic strings as a serious player in the early universe
\cite{Kibble:2004hq}. Whilst it is now clear that cosmic strings cannot play a dominant role
in structure formation their appearance in many scenarios, whether they be motivated by M-theory or supergravity leads us to consider them
as a sub-dominant partner \cite{Contaldi:1998qs,Jeannerot:2003qv,Wyman:2005tu} with 
tensions of the order $G\mu\leq 10^{-6}$. 
Depending on the model and the allowed decay processes for the string network, the bounds on $G\mu$ can be significantly strengthened. These could include the production of
dilatons \cite{Babichev:2005qd}, of gravitational waves \cite{Damour:1996pv} or even the possibility of loops of string being trapped in the compact dimensions leading to a monopole type problem for the strings \cite{Avgoustidis:2005vm}.

For many years the main connection between cosmic strings and superstrings
was that both were much studied but never observed. Indeed, the high tension of fundamental
strings, placed at around the Planck scale, ruled then out as cosmic string candidates.
However, over the past decade or so, more interesting solutions have begun to emerge in string theory, involving warped compactifications or large extra dimensions. One of the by-products has been a realization that these warp factors play
a crucial role in determining energy scales of compactifications \cite{Randall:1999ee}.
In string theory these warp factors are sourced by brane-fluxes and much has been made
of their use in model building\cite{Giddings:2001yu,Kachru:2003aw}. 

\begin{figure}[t]
\includegraphics[scale=0.5,angle=0]{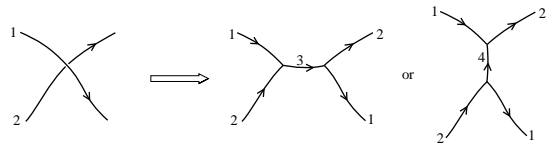}
\caption{Example of formation of string segments due to non-intercommuting .}
\label{nonab}
\end{figure}

Recently the possibility that the same basic mechanism could provide both the inflaton and lead to the formation of  cosmic strings has received renewed attention in the context of braneworlds \cite{bab,tye,Kachru:2003sx,DKVP,DVnew,Copeland:2003bj}. Some consistency conditions for superstrings to act as cosmic strings were given in
\cite{Copeland:2003bj}, in particular the KKLMMT\cite{Kachru:2003sx} picture of IIB
string theory has a warp factor which allows the Dirichlet-fundamental strings
$(p,q)$ bound state to have reduced tensions and behave as cosmic strings. Such cosmic
strings are different to the standard Abelian strings
as they offer more ways of interacting;
two different $(p,q)$ strings could pass though each other, or they could reconnect
in one of the two ways depicted in Fig. \ref{nonab}\cite{Jackson:2004zg}. A network
formed from such strings will therefore have quite different properties. Networks
formed from Abelian strings consist of loops and long strings, $(p,q)$ networks
also contain loops and long string, but there are also {\it links} which start and
end at a three-point vertex Fig. \ref{nonab}. The presence of these links in non-Abelian
string networks has lead authors to speculate the possible existence of a frozen
network which comes to dominate the matter
content of the Universe\cite{bpsnet,Spergel:1996ai,Bucher:1998mh}. A number of authors have begun to reconsider the evolution of the more complicated dynamics associated with cosmic superstrings, and argued in general that scaling solutions are achieved\cite{maria,martins,shellard}. For example, recently
there has been work suggesting that such networks of $(p,q)$ strings reach a scaling behaviour, where the density of strings scales with that of the background energy density of the Universe \cite{Tye:2005fn}.
Here we re-consider the field theory model of \cite{Spergel:1996ai}, which can be thought
of as containing $N$ basic strings and $\half N(N-3)$ bound states to give a total
of $\half N(N-1)$ different species of string. We shall see from the numerical simulations
that as $N$ increases, so does the scaling density. This is to be expected because the
network loses string via the formation of loops, and loops can only form from the reconnection
event caused by two strings of the same type intersecting. Because the chance of a string
meeting the same type decreases for large $N$ then the probability of loop formation
at reconnection goes down.
Although we do not claim this model accurately describes
a network of $(p,q)$ strings it does have the important property that 
strings reconnect as they pass through each other, so creating the links which are
the main difference to the standard cosmic string scenario. An important difference
between our model and that of the $(p,q)$ relates to the allowed string tension. While
our field theory gives all strings the same tension, the tension of a $(p,q)$
string increases for larger $p$, $q$ having the effect that low $p$, $q$
will be preferred dynamically. From the perspective of the field theory model
this would correspond to having $N$ small.

Before discussing the dynamics associated with a network of non-abelian strings, we first remind the reader about some of the basic properties associated with ordinary abelian strings. For a comprehensive review see \cite{vilenkin:1994}.

The initial distribution of a network of strings is comprised of a combination of infinite strings that stretch across the universe and a scale invariant distribution of loops of string. Roughly 80\% of the string is initially infinite in length and 20\% is in loops. There exist a series of characteristic length scales on the network, but the one we are interested here is $\xi$, which corresponds to the typical separation between strings, and is given in terms of the volume $V$ and total length $L$ of strings by $\xi^2 = V/L$. Other possible scales correspond to the typical curvature scale on the strings, and the small scale structure on the string network. As the universe expands, the network evolves and the characteristic length scales change with time as well. There are a number of physical processes that occur allowing the network to lose energy. The long strings can cross themselves forming loops which are chopped off the long strings, strings oscillate and because they are so massive they radiate energy through gravitational radiation. For our case, where we are considering a field theory model, the prominent decay mode will be through particle production. Meanwhile the long strings are stretched by the expansion, which increases the energy stored in the network. Hence there are competing effects which act against each other. Determining precisely how such a network evolves is a challenge that has proved difficult to solve exactly. The equations of motion are non-linear and the requirement of keeping track of the whole network in case new loops are formed from string intersections, requires considerable computing power. The net result is that the correlation length $\xi$ scales with the Hubble radius, leading to the ratio of the string energy density and the background radiation/matter energy density  becoming a constant, implying that the strings provide a fixed contribution to the total energy density in the universe. This scaling solution formed the backbone of investigations into string dynamics. The initial loop distribution is modified to one where the number density of loops of size $l$ falls off as a power law in both the radiation and matter dominated eras, whilst maintaining the scale invariant distribution. The case for non-abelian strings is not as well understood, not least because much less effort has gone into following the evolution of these more complicated objects. Aryal et al \cite{Aryal:1986cp} investigated the formation of $Z_3$ strings. Based on numerical simulations they argued that unlike the case for Abelian strings, the initial loop distribution was not a scale invariant power law distribution but rather it was better approximated by an exponential distribution with the number density falling off exponentially with loop length. The majority of the string length is initially to be found in one network taking up 93\% of the total string length, with the rest in loops and much smaller networks. It raised the important question, what happens to such a non-abelian network, and this was partly addressed in Refs.~ \cite{Vachaspati:1986cc,Spergel:1996ai,McG}? In this paper we will look more closely at the evolution of non-abelan strings, and show that they always appear to enter a scaling regime with a corresponding loop distribution that, rather than having a scale invariant power-law distribution, instead appears to be exponential. We will follow and modify the approach developed by Vachaspati and Vilenkin in  Ref.~\cite{Vachaspati:1986cc} where the authors investigated the evolution of a network of strings connected to monopoles. 

\section{one scale model}

In this section we shall describe an analytic approach to understanding the
scaling regime of a network of non-Abelian string. In what follows we will assume that the
network evolution is described by a single length-scale, $\xi$, and 
shall see that the solution to this model is that a network evolves in a
self-similar fashion with $\xi\sim t$, just as for Abelian networks. We start by
writing down the evolution equation for a network containing a single string species
evolving in an expanding background with Hubble parameter $H=\dot a/a$, where $a$ is the scale factor and $\dot a \equiv \frac{da}{dt}$.
Taking the pressure of the string network as $p=\gamma \rho$, where $\rho$ is the energy density of the strings, and $\gamma$ is the equation of state parameter, then the fluid equation is
\be
 \dot{\rho} = - 3H(1+\gamma)\rho - c_p\rho/\xi.
 \label{evolve-no-loop}
\ee
The first term on the right hand side includes the work done by the system per unit volume per unit time, and the 
second term describes the process of energy loss from strings
by particle emission \cite{Vincent:1997cx}. Taking $\mu$
to be the string tension we define the network length-scale $\xi$ through the string density, 
\be
\rho \sim {\mu \xi \over \xi^3}.
\label{scalingRho}
\ee 
For a scale factor $a\propto t^\nu$, then

\ba
\dot \xi&=&\frac{3}{2}(1+\gamma)\nu \xi/t+\frac{c_p}{2}\\
      &=&\kappa \xi/t+\lambda
\ea
where $\kappa=\frac{3}{2}(1+\gamma)\nu$, $\lambda=\frac{c_p}{2}$. 
This can be solved to give \cite{Vachaspati:1986cc}
\ba
\xi(t)&=&\frac{\lambda}{1-\kappa}t+K t^{\kappa}.
\ea
Causal evolution of the network requires $\xi$ to grow no faster than $t$ so we must have
$\kappa\leq 1$, leading to $\xi\propto t$ asymptotically, typical of self-similar evolution.

Now we include loop production and reconnection. Following \cite{Vachaspati:1986cc}, normally we would expect that most of the loops chopped off the network are likely to be recaptured,  both events occurring on a time-scale of order $\xi(t)$.  However, including expansion, the two events are not quite equal. In a time interval $\xi$, the 
density of strings is reduced by a fraction $\xi/t$, leading to a similar reduction in the probability of loop 
absorption. In other words there is an imbalance with the infinite network losing energy to loops at the rate 
$\dot{\rho}_{\rm loops} \sim - (\xi/t)(\mu \xi/\xi^4)$ or 
\be
\dot{\rho}_{\rm loops} \sim - c_l\rho/t,
\ee
where we expect the constant $c_l\sim 1$. Things change even more in our particular case.  We shall see later that there is not one type of string
but $\half N(N-1)$, which can be thought of as bound states of $N$ types of underlying string.
Now, imagine two of these strings $N_1$ and 
$N_2$ which cross one another. Only when $N_1$ and $N_2$ are the same will they form loops. Hence the 
probability of forming loops is decreased in our case by an additional factor of $2/(\half N(N-1)+1)$. 
We see this by noting that for $n(=\half N(N-1))$ string species there are $\half n(n+1)$ distinct pairings
of string, of which $n$ will pair up the same type. So the chance of a random pairing containing two
of the same species is \mbox{${n \over \half n(n+1)}={2 \over \half N(N-1)+1}$}
In other words we have  
\be
\dot{\rho}_{\rm loops} \sim - {2c_l\over \half N(N-1)+1} {\rho \over t}.
\label{nloops}
\ee
In the rate 
equation (\ref{evolve-no-loop}) we now have
\ba
 \dot{\rho} &=& - 3H(1+\gamma)\rho - c_p\rho/\xi + \dot{\rho}_{\rm loops} \\
&=&- 2\kappa\rho/t - c_p\rho/\xi  - {2c_l\over \half N(N-1)+1} {\rho \over t}.
 \label{evolve-with-loop}
 \ea
Note the form of the loop term is simply to re-normalise the work-done contribution. In other words the solution 
for the characteristic length scale is as we quoted earlier but with a new $\kappa$:
\be
\xi(t) = {\lambda \over 1-\kappa_N}t + K_N t^{\kappa_N}
\label{dsoln-withloop}
\ee
where $\kappa_N = \kappa + \frac{c_l}{\half N(N-1)+1}$.  
Now we have obtained the characteristic length of the network as a function of N we can compare
this to our numerical solutions, which we now describe.

\section{numerical model}
\begin{figure}[ht] 
   \centering
   \includegraphics[width=3in]{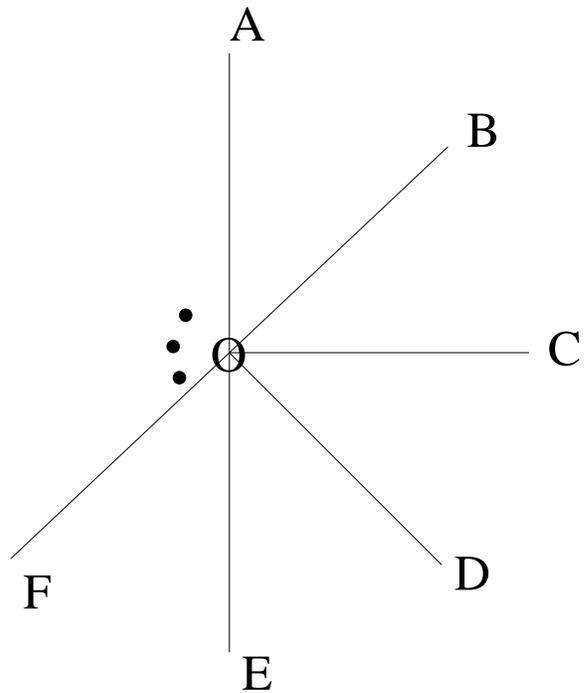} 
   \caption{Vacuum manifold for the sigma model, the ends of the spokes are identified.
   \label{vacuum}}
\end{figure}

\begin{figure}[ht] 
   \centering
   \includegraphics[width=3in]{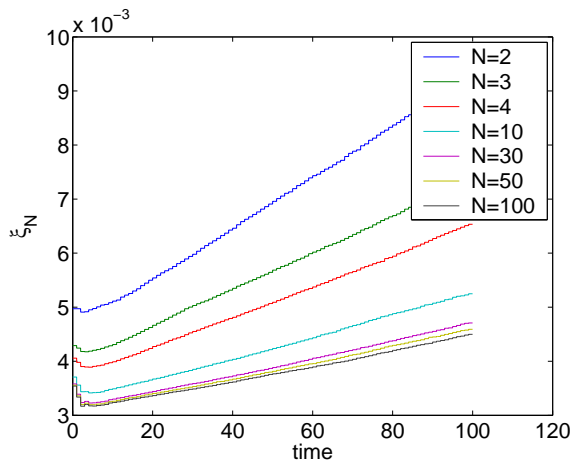} 
   \caption{The evolution of the typical string length-scale as a function of N.
   \label{xiN}}
\end{figure}

\begin{figure}[ht] 
   \centering
   \includegraphics[width=3in]{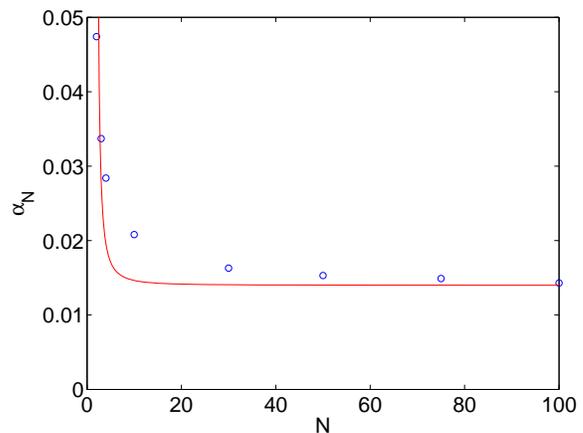} 
   \caption{The slope of $\xi(t,N)$ as a function of N.
   \label{alphaN}}
\end{figure}

In order to model a network of strings which do not intercommute, but rather form links between the two
string segments as they cross (see Fig. \ref{nonab}) we followed the techniques described in \cite{Spergel:1996ai}
for setting up a linear sigma model. This is an approximation for describing a scalar field in the long
wavelength limit where the field is taken to live on its vacuum manifold. In order to have different types
of string this manifold must have more than one non-contractible loop, with each such loop corresponding to
a string solution. We take the vacuum as
given in \cite{Spergel:1996ai}, and depicted in Fig. \ref{vacuum}, which takes the form of $N$ spokes whose
ends are identified, noting that $N=2$ corresponds to the usual case of a single species of string
\cite{footnote}. In general there are $\half N(N-1)$ species corresponding to the number of incontractible loops in the
vacuum. For example, referring to Fig. \ref{vacuum} we could form strings from the loops AOBA, AOCA, BOCB,...
Therefore we evolve a scalar field which takes values in $[0,\pi)\times\bb{Z}_N$
and has a canonical kinetic term. The initial conditions are chosen such
that the field takes random values
at each lattice site and we use a $300^3$ lattice.
The first set of simulations were performed in Minkowski space and the results for
the network length-scale are given in Fig. \ref{xiN}, showing how $\xi$ evolves in time for various $N$.
Note that the network approaches a scaling solution in each case, with $\xi\propto t$. Also, in the simulations
with a larger number of species the network grows more slowly in time, as expected from (\ref{dsoln-withloop}),
corresponding to a higher string density.
The rate of increase of the network length-scale may now be extracted from the simulations 
by writing the late time behaviour of the simulation length-scale as $\xi_{sim}=\alpha t$, which can then
compared to the asymptotic analytic form (\ref{dsoln-withloop})
\ba
\xi(t)\rightarrow{\lambda \over 1-\kappa_N}t
\ea
In Fig.~\ref{alphaN}, we plot $\alpha$ as a function of $N$ for the simulations
given in Fig. \ref{xiN} with the analytic form shown on the graph for comparison. Although the match is by no
means perfect it does illustrate the general behaviour that $\xi$ increases more slowly as the number of string
species increases, and also that the gradient $\alpha$ asymptotes to a constant rather than zero, meaning that
even for an arbitrarily large number of species the network will still scale rather than freeze.
It is worth noting that our model underestimates the value of $\dot \xi_N$, corresponding to an
overestimate of the scaling density. This means that the network has another mechanism by which it can lose
energy. One possibility, identified by \cite{Tye:2005fn}, is that the network will lose energy from the
binding energy of two string species as they merge into a bound state, this could account 
for our overestimate of the scaling density.

\begin{figure}[ht] 
   \centering
   \includegraphics[width=3in]{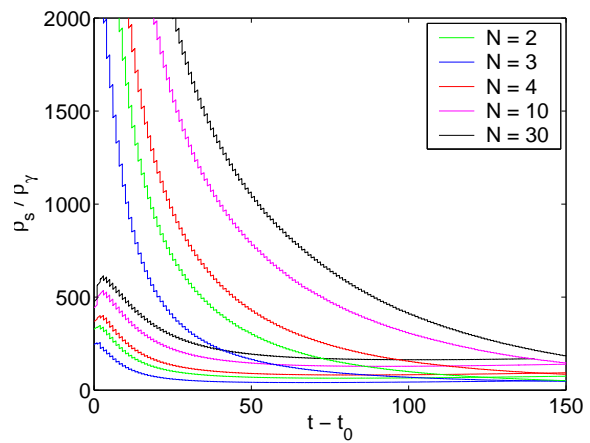} 
   \caption{Scaling in radiation as a function of N.
   \label{radScal}}
\end{figure}

We have repeated the simulations for a radiation dominated Universe and present the results in Fig. \ref{radScal}.
Plotted here is the ratio of the energy density in the string network to that of the background radiation
as a function of time. For each value of $N$ there are two curves shown, each with a different initial value
for the ratio of string density to radiation density.
The figure shows that for a given $N$
each curve seems to approach the same constant value, indicating that there is
a unique scaling solution for each $N$. Unfortunately the simulations run out of dynamic range before we are
able to get accurate values for the scaling density, however we can see from the plots that the scaling density
increases with $N$. As the density in radiation scales as $a^{-4}$, with $a\sim\sqrt{t}$ we have that the
network density during scaling varies as $\rho_s\sim 1/\xi^2\sim 1/t^2$ which gives us the length-scale increasing
as $\xi\sim \alpha t$ as before. Moreover, the simulations in the radiation background indicate that the scaling
density increases with $N$ implying that $\alpha$ reduces as $N$ increases, consistent with our analytic approach.

\begin{figure}[ht] 
   \centering
   \includegraphics[width=3in]{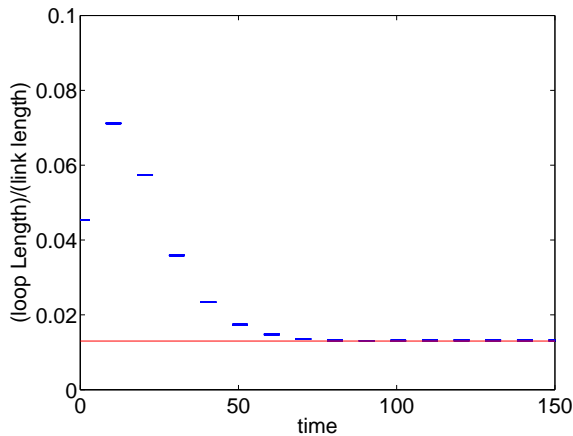} 
   \caption{The ratio of loop length to link length for $N=3$.
   \label{loopLinkRatio}}
\end{figure}

We can provide further evidence for a scaling solution by measuring the total length of string contained
in loops and the total length in links. In a scaling solution there should only be a single length-scale
describing the properties of the network, as such we would expect the ratio of the amount of string in loops
to the amount of string in links to approach a constant. To test this we focused on the case of $N=3$
and ran 100 simulations in order to minimize statistical errors.
The results are plotted in Fig. \ref{loopLinkRatio} where we see how the ratio of loop length to link length
evolves in time, we can clearly see that the ratio tends to a constant, furnishing us with further evidence
that the network approaches scaling. 
We now proceed to describe the relative length distributions of the  loops and links in these non Abelian networks.

\section{length distribution}

\begin{figure}[ht] 
   \centering
   \includegraphics[width=3in]{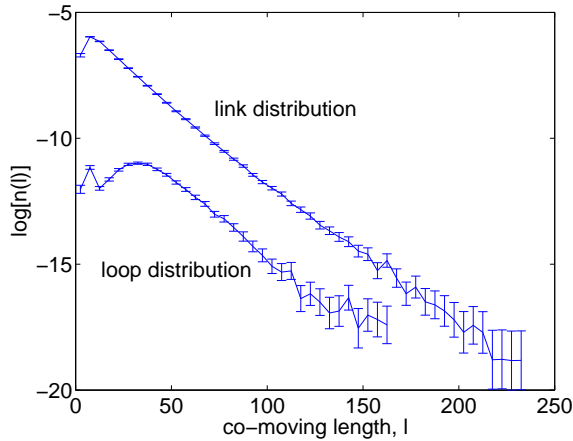} 
   \caption{The distribution of loop/link lengths for $N=3$
   \label{loopLinkDist}}
\end{figure}

We have presented evidence that the string networks enter a scaling regime which can be described by
a single length-scale. This however does not mean that all the loops and links have the same length given
by $\xi$, so we now turn our attention to the distribution of the lengths in loops and links. 
It is well known that the more usual Abelian cosmic string has a power law distribution of loop length,
$n(l){\rm d}l\sim l^{-5/2}{\rm d}l$ \cite{vilenkin:1994}, however our case for multiple species of string
is more like the situation studied in \cite{Vachaspati:1986cc} with three string species that could connect
via heavy monopoles. There it was found that the distribution was more closely modelled by an exponential.
Exponential distributions for string lengths were also found in \cite{Aryal:1986cp} where the statistics
of a random distribution of $\bb{Z}_3$ strings were studied.
Here, we have focused on the case $N=3$ and performed 100 simulations in order to get good statistics for the length distribution of loops and links. The results are given in Fig. \ref{loopLinkDist}, where we have plotted $\ln n(l)$ 
{\it vs} $l$, when the system has approached the scaling regime. 
At large $l$ the  nearly-straight line of the
plot clearly shows that the distributions are more accurately described by exponentials rather than
power laws.
We may model this by an exponential, dimensionless loop production function $f(l/\xi)$
by adapting the arguments found in \cite{vilenkin:1994} for the Abelian string. 
This function $f(l/\xi)$ describes the loss from the network of strings of length $l$.
We consider the network
to be in the scaling regime described by the length-scale $\xi$ such that (\ref{scalingRho}) holds for 
the network. The loop production function $f(l/\xi)$ is defined such that 
\mbox{$\mu f(l/\xi){\rm d}l/l$} is the energy lost in loops of length between $l$ and $l+{\rm d}l$
in volume $\xi^3$ per unit time. This then means that the loop density evolves according to
\ba
\dot\rho_L&=&-3H\rho_L+g\mu f(l/\xi)/\xi^4,
\ea
where $g$ is a Lorentz factor to account for the loops created with
non-zero centre of mass kinetic energy. 
We may rewrite this expression in terms of the dimensionless
parameter $x=l/\xi$ and integrate to find
\ba
\rho_L(l,t)=\frac{g\mu l^{3\nu-3}}{\alpha^{3\nu+1}t^{3\nu}}\int_{l/\alpha t}^\infty {\rm d}x\;x^{2-3\nu} f(x).
\label{analyticDist}
\ea
Here we have taken $a(t)\sim t^\nu$, $\xi=\alpha t$, note that the radiation era corresponds to
$\nu=\half$. In order to get an exponential distribution of loop lengths we must therefore have a loop
production function which contains an exponential, for example we may have
\ba
f(x)=b x^p\exp(-\beta x),
\ea
where $b,\,p$ and $\beta$ are constants.  
Taking this form and using $n(l,t)=\rho_L(l,t)/l$, (\ref{analyticDist}) gives: 
\ba\nonumber
n(l,t)&\simeq& \frac{bg\mu l^{3\nu-4}}{\alpha^{3\nu+1}\beta^{3+p-3\nu}t^{3\nu}}
             \Gamma(3+p-3\nu,\beta l/\alpha t)\\
   &\rightarrow&\frac{bg\mu }{\alpha^{3+p}\beta t^{2+p}}l^{p-2}\exp(-\beta l/\alpha t)
    \label{num-den-predict}
\ea
where in the final expression we have taken the limit $x>>1$. In principle we should be able to determine the 
dependence on $p$ from the numerical simulations, but this is not possible as it stands due to the statistical 
errors in the data at large loop length. In other words we can not yet determine the form of any power law 
correction to the pure exponential distribution. 

We finally return to the distribution of the total string lengths in links and loops in the scaling regime, as shown in Fig.~(\ref{loopLinkRatio}). We can understand the late time behaviour of this Figure  based on our analytic approximations. If we consider the link length, $L_{\rm links}$ to be simply the difference between the total length of string $L$ and the length of string in loops, $L_{\rm loops}$, then we can write :
\ba
{L_{\rm loops} \over L_{\rm links}} = {1 \over (\frac{L}{L_{\rm loops}} -1)}
\label{loop-link-ratio}
\ea
Now in a large volume V, where the string tension is $\mu$ we know that 
\ba
{L \over L_{\rm loops}} = {\rho \over \rho_{\rm loops}}
\label{l-loop-ratio}
\ea
where $\rho_{\rm loops}$ is the energy density in loops defined from Eq.~(\ref{analyticDist}) by
\ba 
\rho_{\rm loops} =  \int_{t_0}^{\alpha t} \rho_L(l,t) dl
\label{def-rho-loop}
\ea
the limits of integration accounting for the largest loops excised at time $t$ and the smallest loops excised at time $t_0$. Performing the integral we find 
\ba 
\rho_{\rm loops} = {b g \mu \over \alpha^3 t^2} {1 \over \beta^{3-3\nu +p}} F(\beta,p,t)
\label{rho-loop}
\ea
where
\ba 
F(\beta,p,t) = \int_{\frac{t_0}{\alpha t}}^1 x^{3(\nu-1)} \Gamma(3(\nu-1) +p,\beta x).
\label{def-F}
\ea
Now, $F$ is independent of $t$ at late times, hence we can extract the time dependence in Eq.~(\ref{rho-loop}) without knowing the full form of $F$. Now given $\rho = \frac{\mu}{\xi^2}$, and $\xi = \alpha t$, we find the late time behaviour in Eq.~(\ref{l-loop-ratio}) 
\ba
{L \over L_{\rm loops}} = {\alpha  \beta^{3-3\nu +p} \over b g F}
\label{l-loop-ratio-final}
\ea
which is independent of time, as required. 

\section{conclusions}

In this paper, in an attempt to model (p,q) strings arising out of compactifications of type IIB string theory, we have studied how a non-Abelian network of strings evolves. Such networks
contain multiple vertices where many different types of string join together. For example, $Z_3$ strings have triple vertices where three different types of string join together. The result of these more complicated vertices  is that  non-Abelian networks have a richer variety of 
re-connections than their counterpart  Abelian networks. Given this rich structure, from a cosmological point of view, 
we need to know whether such networks approach a scaling solution where they track the dominant background energy density or simply freeze, and stretch with the expansion of the universe leading to a string dominated Universe. By re-considering the field theory model of \cite{Spergel:1996ai}, which can be thought of as containing $N$ basic strings and $\half N(N-3)$ bound states to give a total
of $\half N(N-1)$ different species of string, we have shown both numerically and analytically that we always approach a scaling solution for all $N$, in other words the system does not freeze. All that happens is that as $N$ increases, so does the scaling density, asymptoting to a fixed value independent of $N$ for large $N$. Moreover, we have seen firm evidence that the associated distribution of loops and links is not governed by a power law as is the case for Abelain and $Z_2$ strings. Rather, as initially observed on smaller lattices by Aryal et al \cite{Aryal:1986cp} and Vachaspati and Vilenkin  \cite{Vachaspati:1986cc}, the distribution drops off exponentially fast with loop size. We have attempted to explain this distribution based on our one-scale model, and argued why the loop-link ratio tends to a constant at late times for the case of $Z_3$ strings. There is much that remains to be done. Perhaps the most obvious thing is to allow the strings to have different tensions in this field theory model. In a very nice paper, Tye et al have been doing just that in the context of modelling cosmic superstrings \cite{Tye:2005fn}. It would be worthwhile
to see whether the field theory mimics 
their results or not. In particular once we have different tensions, it would
be interesting to see how much string of each tension exists in a
scaling network.
One of the nice features about our approach is that we can produce the strings in an 
initial phase transiton and watch them evolve. It would be interesting to see how quickly
the heavier tension strings appear, starting say from a network of light strings.

\vspace{1cm}
\noindent
{\large\bf Acknowledgements} 
The authors would like to thank PPARC for financial support, Nuno Antunes for
collaboration in the early stages of the project and Mark Hindmarsh for
useful discussions. P. M. S. would like to acknowledge extensive use of the UK
National Cosmology Supercomputer funded by PPARC, HEFCE and Silicon Graphics.


\begin{thebibliography}{99}
\bibitem{footnote} Strictly speaking the vacuum does not form a manifold as each point is not homeomorphic to $\bb{R}$.
\end{thebibliography}
\end{document}